**Article type: Communication**

**Title** Strain-mediated high conductivity in ultrathin antiferromagnetic metallic nitrides

*Qiao Jin, Hu Cheng, Zhiwen Wang, Qinghua Zhang, Shan Lin, Manuel A. Roldan, Jiali Zhao, Jia-Ou Wang, Shuang Chen, Meng He, Chen Ge, Can Wang, Hui-Bin Lu, Haizhong Guo, Lin Gu, Xin Tong, Tao Zhu, Shanmin Wang,* Hongxin Yang,* Kui-juan Jin,* and Er-Jia Guo**

Mr. Q. Jin, Dr. Q. H. Zhang, Miss. S. Lin, Dr. J. Zhao, Mr. S. Chen, Miss. M. He, Dr. C. Ge, Prof. C. Wang, Prof. H. B. Lu, Prof. T. Zhu, Prof. L. Gu, Prof. K. J. Jin, and Prof. E. J. Guo
Beijing National Laboratory for Condensed Matter Physics and Institute of Physics, Chinese Academy of Sciences, Beijing 100190, China
E-mail: kjjin@iphy.ac.cn and ejguo@iphy.ac.cn

Dr. H. Cheng and Prof. S. M. Wang
Department of Physics, Southern University of Science and Technology, Shenzhen, Guangdong 518055, China
E-mail: wangsm@sustech.edu.cn

Mr. Z. Wang and Prof. H. X. Yang
Ningbo Institute of Materials Technology & Engineering, Chinese Academy of Sciences, Ningbo 315201, China
E-mail: hongxin.yang@nimte.ac.cn

Dr. M. A. Roldan
Eyring Materials Center, Arizona State University, Tempe, AZ 85287, United States

Dr. J. Zhao and Prof. J. O. Wang
Institute of High Energy Physics, Chinese Academy of Sciences, Beijing 100049, China

Mr. S. Chen and Prof. H. Z. Guo
School of Physical Engineering, Zhengzhou University, Zhengzhou 450001, China

Prof. Xin Tong and Prof. T. Zhu
China Spallation Neutron Source, Institute of High Energy Physics, Chinese Academy of Sciences, Beijing 10049, China

Prof. C. Wang, Prof. T. Zhu, Prof. L. Gu, Prof. K. J. Jin, and Prof. E. J. Guo
Songshan Lake Materials Laboratory, Dongguan, Guangdong 523808, China
and School of Physical Sciences, University of Chinese Academy of Sciences, Beijing 100190, China

Prof. E. J. Guo
Center of Materials Science and Optoelectronics Engineering, University of Chinese Academy of Sciences, Beijing 100049, China

**Abstract**: Strain engineering provides the ability to control the ground states and associated phase transition in the epitaxial films. However, the systematic study of intrinsic characters and their strain dependency in transition-metal nitrides remains challenging due to the difficulty in fabricating the stoichiometric and high-quality films. Here we report the observation of electronic state transition in highly crystalline antiferromagnetic CrN films with strain and reduced dimensionality. Shrinking the film thickness to a critical value of ~ 30 unit cells, a profound conductivity reduction accompanied by unexpected volume expansion is observed in CrN films. The electrical conductivity is observed surprisingly when the CrN layer as thin as single unit cell thick, which is far below the critical thickness of most metallic films. We found that the metallicity of an ultrathin CrN film recovers from an insulating behavior upon the removal of as-grown strain by fabrication of first-ever freestanding nitride films. Both first-principles calculations and linear dichroism measurements reveal that the strain-mediated orbital splitting effectively customizes the relatively small bandgap at the Fermi level, leading to exotic phase transition in CrN. The ability to achieve highly conductive nitride ultrathin films by harness strain-controlling over competing phases can be used for utilizing their exceptional characteristics.

**Main text**

Ultrathin conductive films hold tremendous technological interests in the transparent displays, flexible electronics, photovoltaics, and *etc*. The industrical applications require that the ultrathin layers possess excellent mechanical properties, exceptional-well thermal stability, and corrosion resistance.[1,2] Transition metal nitrides (TMNs) is a large group of materials that typically have high conductivity and chemical stability to meet these requirements. Among them, chromium nitride (CrN) is an exceptional-well prototype material combined high conductivity with an unique antiferromagnetic (AF) configuration.[3, 4] The natural fully compensated magnetic moment in CrN makes it absence of parasitic stray fields and be robust against the magnetic perturbations, leading it is ideal for the secure data storage and memory devices.[5-10] Another advantage against the conventional metallic AF materials containing the noble metals such as MnPt, [11] IrMn, [12] and $Mn_2Au$, [13] the CrN is an economic and stable AF material, which is central requirements for developing industrial products with combined superior characteristics.

Typically, the conductivity of a metallic film changes with strain and decreases as reducing the dimensionality. The strongly competition between the itinerant and correlated electrons gives rise to the observed thickness-driven electronic state transition. In the transition metal oxides (TMOs), a metallic layer becomes an insulator as the film thickness reduces below 4–5 unit cells (u.c.). [14-18] However, the systematical studies of thickness- and strain-driven electronic phase transition on the TMNs are still lacking, owing to the difficulty in fabricating the stoichiometric and high-quality thin films, especially CrN. At ambient conditions, bulk CrN has a cubic rock-salt paramagnetic structure and it transforms into an orthorhombic AF structure below the Néel temperature ($T_N$) of ~ 280 K.[19] Coupling with the magnetostructural transition at $T_N$, an electronic transition occurs simultaneously. However, the question of whether intrinsic CrN is metallic or insulating is still controversial and the room-temperature resistivity value ranges over three orders of magnitude.[20-26] Some groups demonstrated that CrN presents a semiconductor-to-insulator transition across $T_N$;[20-22] while others argued that CrN maintains the metallic state at all temperatures.[23-26] These controversial results would be closely associated with the crystalline defects such as the lattice disorder and nitrogen deficiency.[20-22] Therefore, it is urgently required to prepare the high-quality single-crystalline CrN for investigating its intrinsic physical properties.

Meanwhile, similar to the TMOs, the lattice distortion in CrN may play a crucial role in the on-site Coulomb electronic correlation and band structure formation. Recent works from Alling and co-workers demonstrate the strong spin-lattice coupling induces magnetostructural transition in CrN,[27-29] suggesting the epitaxial strain would be an effective tuning knob to understand underlying mechanism including the quantum confinement and orbital physics. Here we report the impact of epitaxial strain on structural and electronic state transitions in highly crystalline CrN ultrathin films. An electronic phase transition has been identified in the CrN films below a critical thickness of ~ 30 unit cells (u.c.). The conductivity in CrN layer maintains when its thickness reduces to one unit cell thick. Combined with first-principles calculations, we find that the band splitting and orbital polarization can be effectively tuned by strain, giving rise to the nonlinear variation of strain effect on the electrical conductivity.

Previously, CrN films had been prepared by reactive magnetron sputtering [26, 30] and RF nitrogen plasma-assisted MBE. [31] The deposition of CrN films by pulsed laser deposition (PLD) technique is rare and challenging. The presence of nitrogen vacancies is unavoidable using a plasma source due to the inactive Cr metal and inert N ions. The specimens investigated in this work were prepared by laser ablating from a high-pressure synthesized stoichiometric CrN target (see Experimental Section and Figure S1, Supporting Information). [32-34] The structural and compositional characterizations of CrN films are summarized in **Figure 1**. As shown in Figure 1b, the representative atomic-resolved cross-sectional high-angle annular dark-field (HAADF) image of a 4-u.c.-thick CrN film grown on a MgO substrate were taken by scanning transmission electron microscopy (STEM). The results confirm the *c*-orientation and the coherent growth of the CrN film. The spatial resolved electron-loss energy spectroscopy (EELS) maps for the O and N *K*-edges indicate the uniformed distribution of elements and the atomically sharp interface with minimum chemical intermixing (Figures S2 and S3, Supporting Information). X-ray diffraction (XRD) measurements show that the as-grown 50-u.c.-thick CrN film exhibits clear Kiessig fringes up to nine orders and a narrow rocking curve at the CrN 002 reflection with full width at half maximum (FWHM) of ~ $0.04^{o}$, which is comparable with that of MgO substrate (Figure 1c). The CrN films with a thickness below 50 u.c. were coherently grown on MgO substrates (Figure 1d). The lattice constants of the CrN films are relaxed towards its bulk form ($a = c =$ 4.15 Å) as increasing film thickness (Figure S4, Supporting Information). Typical root-mean-square (rms) roughness of ~ (4 ± 1) Å for the CrN/MgO interfaces and CrN surfaces was obtained by fitting x-ray reflectivities of all samples (Figure S5, Supporting Information). The AF nature of CrN films is confirmed by measuring the field-dependence of magnetization (Figure S6, Supporting Information). Both magnetic and electrical measurements identify the first-order phase transition across the $T_N$. We performed the polarized neutron reflectivity (PNR) measurements on a $TaO_x$/Ta/Co/CrN sample to identify the magnetic origin under external fields. As shown in Figure 1e, the large splitting between the reflectivities from spin-up and spin-down neutrons indicates the magnetic contribution along the in-plane direction. Figure 1f shows the chemical (*n*SLD) and magnetization (*m*SLD) profiles of this sample. The magnetic moment originates from the Co layers solely, confirming the AF CrN does not contribute to the total magnetization. Previous work highlights the profound influence of stoichiometry on the electronic properties of CrN. [30] We performed the room-temperature elemental-specific EELS (Figure S2, Supporting Information) and XAS (Figure S7, Supporting Information) measurements on the as-grown CrN films. The results reveal that the valence state of Cr is nearly 3+ and the Cr : N molar ratio is close to 1, suggesting the a sufficiently low concentration of nitrogen vacancy, which is beyond the detection limit of those techniques.

To explore the electronic properties of CrN films, we performed transport measurements on the CrN films with various film thicknesses. **Figure 2**a shows the $\rho_{xx}$-$T$ curves for the CrN films with film thickness ($t$) ranging from 1 to 500 u.c. Surprisingly, it is found that the conductivity of CrN films can be retained as its thickness decreases down to 1 u.c. with a small resistivity of ~ 1 Ω·cm at ambient conditions. In fact, this sub-nanometer thickness is well below the typical critical thickness (4−6 u.c., ~ 2 nm) of most correlated metallic oxide films.[14-18] This surprising result may be associated with a small amount of oxygen contaminations at the surface of the CrN monolayer. Recent works demonstrate that the substitutional replacement of N with O atoms at the CrN surface may act as *n*-type dopants that yield localized states near the conduction band, leading to an enhanced conductivity. [35, 36] $T_N$ is 253 K for a 500-u.c.-thick CrN film and $T_N$ drastically reduces as decreasing the film thickness (Figure 2b). We find $T_N$ shows a finite-size scaling relation, which is proportional to the $B(t)^{\lambda}$, where the extrapolated B = 14 ± 0.5 and the shift exponent for the finite-size scaling factor $\lambda = 0.46 \pm 0.05$, in consistent with the previous reports. [37, 38] Similar phenomena have been observed in other AF thin films, such as CoO, $Mn_3Ir$, and $Cr_2O_3$, [39-42] suggesting a vanish of AFM order in the ultrathin CrN layers. This behavior can be attributed to the limited magnetic correlation length [39] and reduced thermal stability of the AFM domains with decreasing film thickness. [43] Meanwhile, the lattice of an ultrathin CrN film is heavily confined by the substrates, thus the low-temperature structural transition is inhibited and the metamagnetic transition probably vanishes as the thickness reduces below a critical value.

The electronic states of CrN films were determined by calculating the first derivation $d\rho_{xx}/dT$. For $t$ above 30 u.c., CrN films exhibit a metallic phase at high $T$. A continuous increment of $\rho_{xx}$ and the metal-to-insulator transition in CrN films is observed as decreasing thickness. Figure 2c shows the conductivities ($\sigma_{xx}$) of the insulating CrN films as a function of $T^{-1/3}$. A linear relationship of $\sigma_{xx}$ ~ $\exp(T_0/T)^{-1/3}$, where $T_0$ is the temperature normalization constant, is observed. The low temperature transport properties of insulating CrN films can be well-described by a variable-range hopping (VRH) mechanism with involving the electron-electron interaction, [44] indicating that such an electronic transition is associated with a dimensional crossover from a 3D metallic state to a 2D insulating state. Figure 2e shows a linear dependence of $\rho_{xx}$-$t$ at the room temperature when $t$ is in excess of 30 u.c. as expected for the uniform films; however, reducing the sample thickness below a critical value leads to a significant increase of $\rho_{xx}$ by four orders of magnitude. Furthermore, we also measured the Hall mobility ($\mu$) and carrier density ($n$) of CrN films at 300 K (Figure 2f). $n$ keeps ~ $(1-1.5)\times10^{20}$ cm$^{-3}$ when $t$ is above 30 u.c. and reduces dramatically to $1.42\times10^{17}$ cm$^{-3}$ as $t$ reduces to 2 u.c. The dimensional confinement localizes the itinerant electrons, leading to a significantly reduced carrier density. We find the $\mu$ is independent with $t$. For $t$ below 30 u.c., $\mu$ is nearly constant and flatulates between 134 and 147 cm$^2$·V$^{-1}$·s$^{-1}$. With increasing $t$, the moderate tensile strain (~1.45%) releases gradually, the occurrence of small amount of misfit dislocations may play a dominating role in reducing $\mu$. In the thick CrN films, the conduction mode turns into 3D, the dislocation effect is negligible; thus the electron mobility recovers to high values of 180−230 cm$^2$·V$^{-1}$·s$^{-1}$. Note that the electron mobilities in our films are almost two orders of magnitude larger than the best value (*i.e.,* 1.5 cm$^2$·V$^{-1}$·s$^{-1}$) reported in bulk CrN, [30, 45, 46] further confirming the stoichiometry and homogeneity of CrN films. [30] The similar trends of $\rho_{xx}$-$T$ and $\rho_{xx}$-$t$ were also observed in the compressively strained CrN films (Fig. S8, Supporting Information), suggesting a universal rule governing the electronic transition. We calculated the lattice volume ($V_{pc}$) and density of CrN films from XRD measurements (Figure 2d). As $t$ decreases from 500 to 30 u.c., both $V_{pc}$ and density keep nearly constant values, close to the bulk values. Obviously, $V_{pc}$ increases by ~1.2 %, while density concurrently reduces by a factor of ~0.75 when *t* decreases to a few monolayers (Fig. S5, Supporting Information). We hypothesis that such anomaly in the density may attribute to the small amount of nitrogen-to-oxygen substitutions in the CrN surface with a thickness of a few unit cells. [35, 36] This reaction is unavoidable without capping layer at ambient conditions. It is very challenge to quantify the quantity of O content at the CrN surface layers because of the $O_2$ contamination when exposes in air. However, it does not influence our main conclusions on the thickness-driven electronic transition because the affected region is far below our critical film thickness and the change of conductivity is small.

To test this hypothesis, we grew a 20-u.c.-thick CrN film on a $Sr_3Al_2O_6$-capped MgO substrate (**Figure 3**a). The substrate can be readily removed by washing the soluble $Sr_3Al_2O_6$ layer away in water. This method eventually helps to prepare a first-ever freestanding CrN film. [47-49] As shown in Figure 3b, we measured the $\rho_{xx}$-$T$ curves for both strained and freestanding CrN films. Without strain, the freestanding CrN film present a clear transformation from insulating phase into metallic phase with $T_N$ of ~ 200 K. The lower $T_N$ in the freestanding CrN film comparing with bulk CrN may be attributed to the natural folding and local structural defects, likely due to the existence of a strong spin-lattice coupling in CrN. [27] Figure 3c shows the $d_{x^2-y^2}$ and $d_{3z^2-r^2}$ orbitals are degenerated with lowering the energy of $d_{x^2-y^2}$ orbital due to the strain-induced orbital splitting in the tensile-strained CrN film. The electrons will prefer occupying the $d_{3z^2-r^2}$ orbital. However, for the freestanding CrN film, the release of epitaxial strain will remove the degeneracy of $e_g$ band. Therefore, the crystal-field energy will reduce to the level of a bulk CrN. The energy of $d_{x^2-y^2}$ and $d_{3z^2-r^2}$ orbitals will be the same, thus the positions and energies of XAS peak are identical. Moreover, a series of CrN films with thickness of 20-u.c. and 70-u.c. are grown on various substrates (Figures S9-S12, Supporting Information), [50-52] allowing to tune the in-plane strain over a wide range both statically and dynamically (Table S1, Supporting Information). We note that the calculated Poisson ratio of CrN thin films is 0.17-0.19, which is smaller than most of known ceramics.[53-55] We attribute such anomaly to the ultra large bulk modulus (~ 361 GPa) of CrN compared to those of nitrides and oxides.[56] Such high stiffness in CrN would lead to small deformation under the epitaxial strain. As shown in **Figure 4**a, the compressive strain largely reduces the conductivity of CrN films. At room temperature, $\rho_{xx}$ of the 70-u.c.-thick films increases linearly with increasing compressive strain above −1.5%, while $\rho_{xx}$ slightly changes in the strain range between −1.5% to 1.5%. These results demonstrate the transport properties of CrN films strongly depend on the epitaxial strain, which profoundly modifies its electronic structure.

In cubic CrN, each Cr atom is octahedrally coordinated by six anions, leading to the splitting of Cr: $d$ orbitals into the two- and three-fold degenerated $e_g$ and $t_{2g}$ bands, respectively. We expect that the energy levels of majority-spin (or spin-up) states are lower than that of minority-spin (or spin-down) states. $Cr^{3+}$ ions have a high-spin state ($S$ = 3/2) and the $t_{2g}$ orbitals are half-filled, while the $e_g$ orbitals are nominally empty, as shown in the inset of Figure 4c. [56-58] However, competition between the crystal-field energy ($\Delta_{CF}$) and Hund's rule coupling strength will dictate the electron redistribution of the sample. Using x-ray linear dichroism (XLD) (Figure 4b), we further characterized the samples' electronic properties by probing the unoccupied states (holes) of the $d_{x^2-y^2}$ and $d_{3z^2-r^2}$ orbitals. Evidently, the peak energy of the $d_{x^2-y^2}$ orbital is lower than that of the $d_{3z^2-r^2}$ orbital when CrN films under tensile strain. As expected, the situation reverses in the compressive strained CrN. The thus-determined XLD reflects the configuration of Cr: $3d^3$ orbitals. Negative XLD values for tensile-strained CrN films suggest a highly occupied $d_{x^2-y^2}$ orbital compared to that of the $d_{3z^2-r^2}$ orbital. The sign of XLD changes as the strain state of CrN films varies from the tensile to compressive strain. Previous work has demonstrated that the electron correlation in CrN is mainly determined by the Cr-Cr bond distance ($r$) (e.g., $r \approx 2.933$ Å at ambient pressure) and orbital hybridization because the Cr-N bond distance barely changes with pressure (or strain).

[56] Under the tensile strain, the reducing $\Delta_{CF}$ ($\propto r^{-5}$) term will give rise to a decreasing splitting between $t_{2g}$ and $e_g$ orbitals. [59] In this state, the itinerant electrons move freely within conduction band, leading to a metallic phase. On the contrary, this energy term increase, as the in-plane compressive strain is applied. When the $t_{2g}$–$e_g$ orbital splitting within the order of the splitting inside the $t_{2g}$ and $e_g$ manifolds, the bands still cross the Fermi level and the CrN films retain their metallic phase. However, under a large structural distortion, the $t_{2g}$–$e_g$ orbital splitting substantially increases to form a band gap. Thus, high compressive strain will make the sample become a correlated insulator. The same effects were observed in the pressure-induced electronic state transition in bulk CrN. [56] The charge redistribution associated with structural transformation occurs between the $t_{2g}$ and $e_g$ orbitals, leading to an itinerant-to-localized electronic crossover with decreasing bond distance.

To understand the film thickness and strain effects on the electronic states of CrN films, we performed first-principles calculations within the framework of density-functional theory (DFT) as implemented in the Vienna *ab initio* simulation package (VASP).[60, 61] **Figure 5**a shows the evolution band structure of CrN film as a function of thickness. For the CrN films thicker than 10 u.c., there is a hole pocket formed at the M point, meanwhile there is an electron pocket formed at the Γ point, indicating the films are in a semi-metallic phase. For the CrN films with a thickness less than 8 u.c., the hole and electron pockets disappear, and an indirect band gap appears, suggesting the electronic structure of CrN films transforms into an insulating state. The band gap increases as reducing the film thickness. The quantitative analysis of band gaps is obtained by subtracting the energy difference between the values at the bottom of the conduction band and the top of the valence band. Figure 5d shows thickness dependence of the band gap in CrN ultrathin films. The positive (negative) values of band gap correspond to insulating (metallic) state of the CrN films. The calculated critical thickness of ~ 10 u.c., at which the electronic state transition occurs. Recently, Botana *et al.*[57] predicted a critical thickness of 6−10 u.c. (*depending on the crystallographic* *structure*) in the CrN films for closing the band gap. They further strengthen that the metallic state become stable by structural relaxation which tend to favor a gap closure. These conclusions are consistent with ours with increasing film thickness and reducing epitaxial strain. The band gap opening in the AFM structured CrN also agrees with a recent calculation using the LSDA+U approach by Herwadkar *et al.*,[64] suggesting a robust electronic structure transition predicted by various theoretical calculations. The difference in the theoretical calculations and our experimental results (~30 u.c.) may be attributed to the epitaxial strain and measuring temperature well above the absolute zero.

We further calculated the band structures of the 10-u.c.-thick CrN films with and without tensile strain (Figures 5b and 5c). The lineshapes of both band structures are nearly identical with a tiny difference close to the Fermi level. For a CrN film with a tensile strain of 2%, the band gap is ~ 20 meV (Figure 5e). When the tensile strain is released, the band gap closes with a relative gap of −6.1 meV. These results indicate that the electronic state of CrN films undergo an insulator-to-metal transition after the removal of tensile strain, in consistent with our transport measurements on the freestanding CrN films. The strong localization of the $t_{2g}$ orbitals in CrN leads to a highly conductive state observed in the CrN ultrathin films. Therefore, from both first-principles calculations and experimental observations, we confirm that the itinerant-electrons in the highly conductive CrN can be significantly modified by the strain-driven band splitting, leading to the observed phase transition. Please note that our calculations are performed assuming the absolute zero temperature and a C-type AF CrN, which is the most stable spin texture. Similar calculations have also been conducted in the G-type CrN layers with different thickness. The tendency of band gap closing with reduced dimensionality does not change under different magnetic orders. [27, 52]

In summary, we report the observation of structural transformation and electronic phase transition in high-crystalline antiferromagnetic CrN films upon strain and reduced dimensionality. The critical thickness of phase transition in CrN films is about an order of magnitude larger than that of correlated TM oxide films, indicating a stronger electron-electron correlation and a higher on-site Coulomb energy in CrN. We demonstrate that the strain controls the degeneracy in these symmetry-localized orbitals effectively, yielding a dramatic change in the orbital splitting and polarization, further tailoring the electronic states of CrN. Our work demonstrates the importance of interplay between the structural and electronic degrees of freedom in TM nitrides, similar to their oxide counterpart, where the dominant physical properties compete within a narrow energy scale. From the practical application point of view, the conductive CrN films with a thickness of 1 u.c. down to the physical limit suggest it is a potential candidate for developing the transparent conductive coatings used in the photovoltaics, displays, and *etc*. Recent theoretical calculations by Zhang *et al*. [62] shed light on a robust ferromagnetism with a very high Curie temperature of 675 K in the CrN monolayer or bilayers due to *p*-*d* exchange interaction. Our own calculations demonstrate that the monolayer-thick CrN exhibits a strong Dzyaloshinskii-Moriya interaction (DMI) and possesses the stabilized skyrmions in the two-dimensional space.[63] Extensive works have been stimulated to realize these fascinating predictions experimentally. Furthermore, within this work, the creation of highly conductive freestanding transition metal nitride membranes strengthens the great opportunity to combine these technically important materials with other low-dimensional materials towards functional spintronic devices, flexible electronics, and CMOS techniques in future.

**Experimental Section**

**High-quality epitaxial film growth**

The single-crystalline CrN thin films were fabricated by pulsed laser deposition. The film thickness was controlled by counting the number of laser pulses and further confirmed by x-ray reflectivity measurements. The stoichiometric ceramic CrN target was synthesized using a high-pressure reaction route from a mixed $CrCl_3$ and $NaNH_2$ powder. The heating process was performed at 5 GPa and 1200 ºC for 5 min. Then, the recovered powder was sintered as a target at 5 GPa and 1100 ºC for 50 min. These high-pressure experiments were performed at the High-Pressure Lab of Southern University of Science and Technology (SUSTech). Same technique was used for synthesizing the stoichiometric single-crystalline nitrides as described in our previous works. The structural and magnetic phase transition temperature in the ceramic CrN is close to room temperature, confirmed by neutron powder diffraction (NPD) (not shown). The CrN films with a thickness ranging from 500 u.c. down to 1 u.c. were fabricated on both MgO and $SrTiO_3$ substrates (Hefei Kejing Mater. Tech. Co., LTD). The 20-u.c.- and 70-u.c.-thick CrN films were deposited on the $LaAlO_3$, $NdGaO_3$, $SrTiO_3$, $GdScO_3$, PMN-PT, and MgO substrates under the same experimental conditions in order to study the strain effects to their physical properties. During the growth, the substrate temperature and base pressure were kept at 600 ºC and $1\times10^{-8}$ Torr, respectively. The laser frequency and energy density were 3−5 Hz and 1.2−1.5 J/cm$^2$, respectively. All films were cooled down slowly at a rate of −5 ºC/min to room temperature after the film growth. For the magnetotransport measurements, we grew a polycrystalline Co film with a thickness of ~ 3 nm on top of a 200-u.c.-thick CrN film by magnetron sputtering. The Co/CrN hybrid sample was further capped with a 5-nm-thick Ta layer to prevent the oxidization of Co films.

**Structural, electrical, and magnetic characterizations**

Synchrotron XRD $\theta$-2$\theta$ scans were conducted at the beamline 14B1 of the Shanghai Synchrotron Radiation Facility (SSRF). Synchrotron XRD RSM scans were conducted at the beamline 1W1A of the Beijing Synchrotron Radiation Facility (BSRF) Radiation Facility (SSRF). The x-ray reflectivities, rocking curves, and RSMs were examined by an in-house four-circle x-ray diffractometer using a Cu-$K\alpha_1$ source (Bruker D8 Discovery and Rigaku). Cross-sectional TEM specimens were prepared using the standard focused ion beam (FIB) lift-off process and examined using JEM ARM 200CF microscopy at both Institute of Physics, Chinese Academy of Sciences and Eyring Materials Center (EMC) of Arizona State University (ASU). Elemental-specific EELS maps were obtained by integrating the N *K*- and O *K*-edges signals from interested regions after background subtracting. The resistivity measurements were carried out using standard van der Pauw geometry with Indium contacts by PPMS. During the transport measurements, the typical *ac* current of 10 $\mu$A was applied in order to minimize the heating effect. The magnetic properties of the Ta/Co/CrN sample were probed by SQUID under in-plane magnetic fields. The field dependent magnetization measurements were conducted at the temperature ranging from 5 to 300 K after the sample was field-cooled down to 5 K under a field of ± 1 T.

**Polarized neutron reflectivity measurement**

Polarized neutron reflectivity (PNR) measurements on a $TaO_x$(2 nm)/Ta(3 nm)/Co(3 nm)/CrN(120 u.c., ~45 nm) sample were conducted on the Multipurpose Reflectometer (MR) beamline at the Chinese Spallation Neutron Source (CSNS). The sample was cooled down to 10 K under an in-plane magnetic field of 1 T. The specular reflectivities were measured as a function of the wave vector transfer along the film surface normal. $R^{++}$ and $R^{--}$ represent the reflectivities from the spin-up and spin-down polarized neutrons, respectively. PNR data were fitted using GenX software. The magnetization of Co layer is calculated from the magnetic scattering length density (*m*SLD) profile of the sample.

**X-ray spectroscopic measurements**

XAS measurements were conducted at room temperature in the total electron yield (TEY) mode by measuring the sample-to-ground drain current at both N *K*-edges and Cr *L*-edges on the beamline 4B9B of the Beijing Synchrotron Radiation Facility (BSRF). XLD were characterized by changing incidence angle ($\alpha$) of linearly polarized x-ray beam with respect to the sample's surface plane. When the $\alpha$ = 90°, i.e. the polarization of x-ray is parallel to the in-plane direction, the XAS reflects the $d_{x^2-y^2}$ orbital occupancy ($I_{ip} = I_{90^o}$). When the $\alpha$ = 30°, XAS probes the unoccupied states in both $d_{x^2-y^2}$ and $d_{3z^2-r^2}$ orbitals, with $I_{oop} = [I_{90^o} - I_{30^o} \bullet \sin^2 30^o)/\cos^2 30^o]$. For simplifying the calculations, the signal of ($I_{90^o} - I_{30^o}$) can indirectly reflect the orbital asymmetry of $e_g$ band and is roughly proportional to the exact XLD signals.

**Fabrication of freestanding CrN membranes**

The $Sr_3Al_2O_6$ film with a thickness of ~ 10 nm was firstly grown on top of MgO substrate followed by growing a CrN film with a thickness of ~ 20 u.c. The substrate temperature was kept at 600 °C. All films were grown in vacuum with a base pressure of $1\times10^{-8}$ Torr. After the growth, the surface of sample was hard contacted with PDMS to eliminate the air bubbles and wrinkles. Then, the whole sample was immersed into the de-ionized water to dissolve the $Sr_3Al_2O_6$ layer. For transferring the freestanding CrN membranes, we placed the PDMS support on sapphire wafer with CrN membranes facing towards the wafer. The freestanding CrN membranes detached the PDMS support by heating up the whole sample to around 100 °C for 15 mins. The transport measurements on the freestanding CrN membranes were performed using standard van der Pauw geometry with silver contacts by PPMS.

**First-principles calculations**

Our first-principles calculations were performed within the framework of density functional theory (DFT) implemented in Vienna *ab initio* simulation package (VASP). The projected augmented wave (PAW) method was used to describe the ion-electron interaction and the exchange-correlation effects were treated by the generalized gradient approximation (GGA) in the form of Perdew-Burke-Ernzerhof (PBE) functional. A vacuum slab of 15 Å was adopted along the normal direction to avoid the interaction between the neighboring blocks. The cutoff energy for plane-wave expansion is set to 520 eV and the Brillouin Zone (BZ) was meshed by the 11×11×1 Monkhorst-Pack grids. To describe the strong on-site Coulomb interaction (*U*) caused by the localized 3*d* electrons of Cr, we employed the GGA+ *U* with an effective $U = 3$ eV. [29] The choice of $U = 3$ eV is slightly smaller than that of bulk CrN, [57, 64] because the calculated structural parameters together with the experimental values and the density of states is consistent with the available photoemission spectra.[22] In addition, we also performed the calculations using values of *U* larger than 3 eV. The bandgaps are slightly larger, however, the tendency of bandgap closing as increasing film thickness is consistent with the present calculations. We believe the main conclusions of this paper are independent of the exact choice of *U*. [56] The atomic positions are fully relaxed with the force acting on each atom is less than 0.001 eV/Å. The CrN films with the orthorhombic structure were modeled along the (001) direction. The optimized unit cell parameters from a bulk CrN ($a$ = 5.974 Å, $b$ = 2.983 Å, $c$ = 4.146 Å) were chosen as a starting point to create the supercells. An in-plane $\sqrt{2} \times \sqrt{2}$ supercell and the in-plane checkerboard AFM ordering were constructed for the calculations.

**Supporting Information**
Supporting Information is available from the Wiley Online Library or from the author.

**Acknowledgements**

We thank Z. H. Zhu, G. Q. Yu, and E. K. Liu at Institute of Physics, Chinese Academy of Sciences and Y. W. Cao at Ningbo Institute of Materials Technology and Engineering, Chinese Academy of Sciences for valuable discussions. This work was supported by the National Key Basic Research Program of China (Grant No. 2019YFA0308500), the National Natural Science Foundation of China (Grant No. 11974390), the Beijing Nova Program of Science and Technology (Grant No. Z191100001119112), the Beijing Natural Science Foundation (Grant No. 2202060), and the Strategic Priority Research Program (B) of the Chinese Academy of Sciences (Grant No. XDB33030200). E.J.G. is supported by the Hundred Talents Program from Chinese Academy of Sciences. M.R. acknowledges the use of facilities within the Eyring Materials Center at Arizona State University. RSM and XAS experiments were conducted at the beamline 1W1A and 4B9B, respectively, of the Beijing

Synchrotron Radiation Facility (BSRF) of the Institute of High Energy Physics, Chinese Academy of Sciences were conducted via user proposals. The partial x-ray diffraction measurements were conducted at the beamline 14B1 of the Shanghai Synchrotron Radiation Facility (SSRF) via a general user proposal.

**Figure and figure captions**

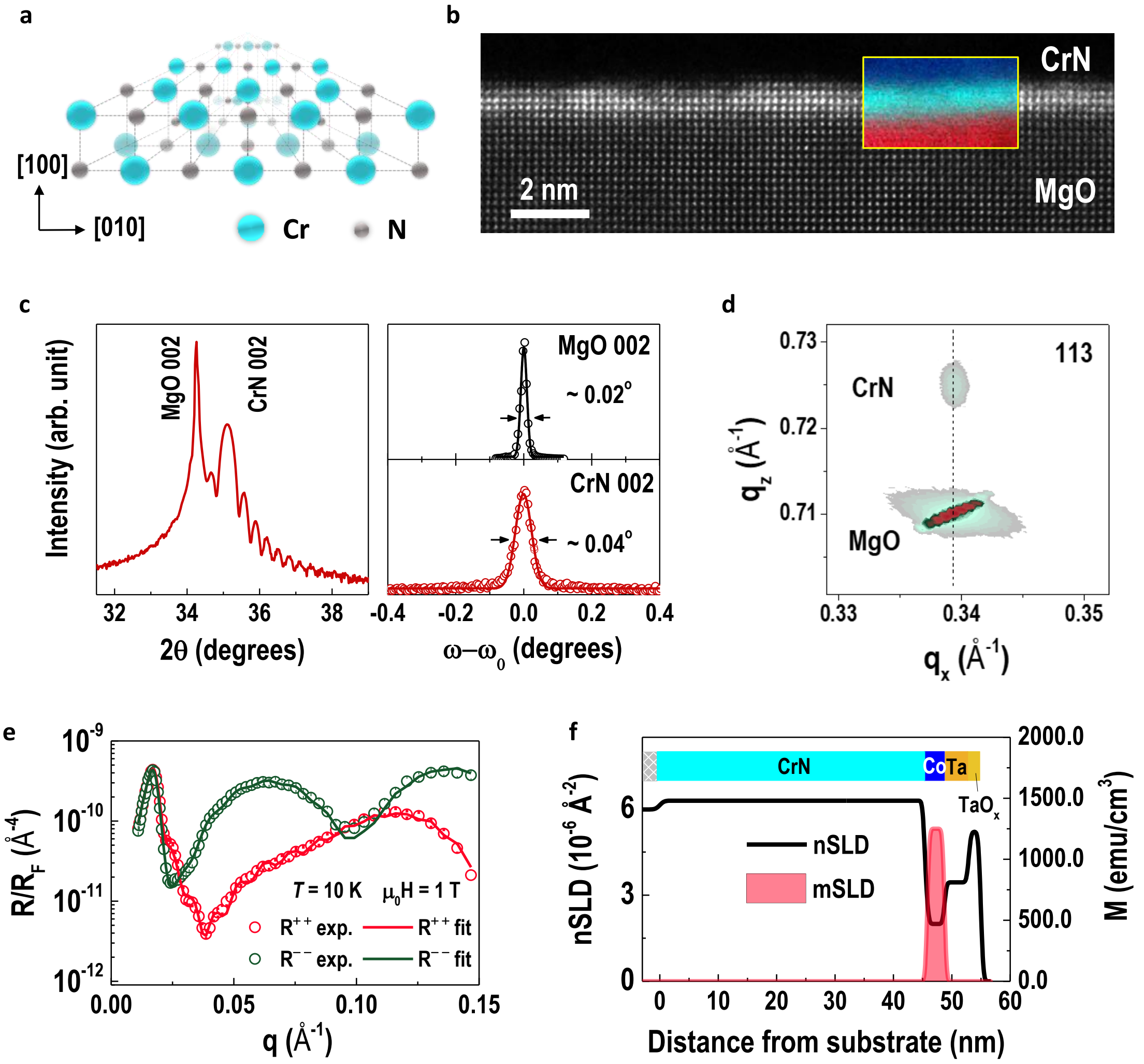

**Figure 1. Structural characterizations of the ultrathin single crystalline CrN films.** (a) Schematic of the CrN lattice structure. (b) Atomic-resolved STEM image of a 4-u.c.-thick CrN film grown on a MgO substrate. The results confirm the *c*-orientation and the coherent growth of the single crystalline CrN film. The inset shows the EELS map from the same area. The spatial resolved electron-loss energy spectroscopy (EELS) maps were taken at the O (red) and N (light blue) *K*-edges, demonstrating the uniformed distribution of elements and the atomically sharp interface with minimum chemical intermixing. (c) XRD $\theta$-$2\theta$ scan of a 50-u.c.-thick CrN film. X-ray diffraction (XRD) measurements show that the as-grown 50-u.c.-thick CrN film exhibits clear Kiessig fringes up to nine orders and a narrow rocking curve at the CrN 002 reflection with full width at half maximum (FWHM) of ~ 0.04°, which is comparable with that of the MgO substrate of ~ 0.02°. (d) RSM of a 50-u.c.-thick CrN film taken around the 113 reflection of the MgO substrates. (e) Reflectivities ($R$) of a TaOx/Ta/Co/CrN sample as a function of wave factor ($q$). $R$ is normalized to the Fresnel reflectivity ($R_F$). Open symbols and solid lines represent the experimental data and theoretical fits, respectively. (f) Nuclear scattering length density (*n*SLD) and magnetization profiles (*m*SLD) of a $TaO_x$/Ta/Co/CrN sample. The magnetization solely comes from the Co layer, further confirming the AF nature of CrN thin films.

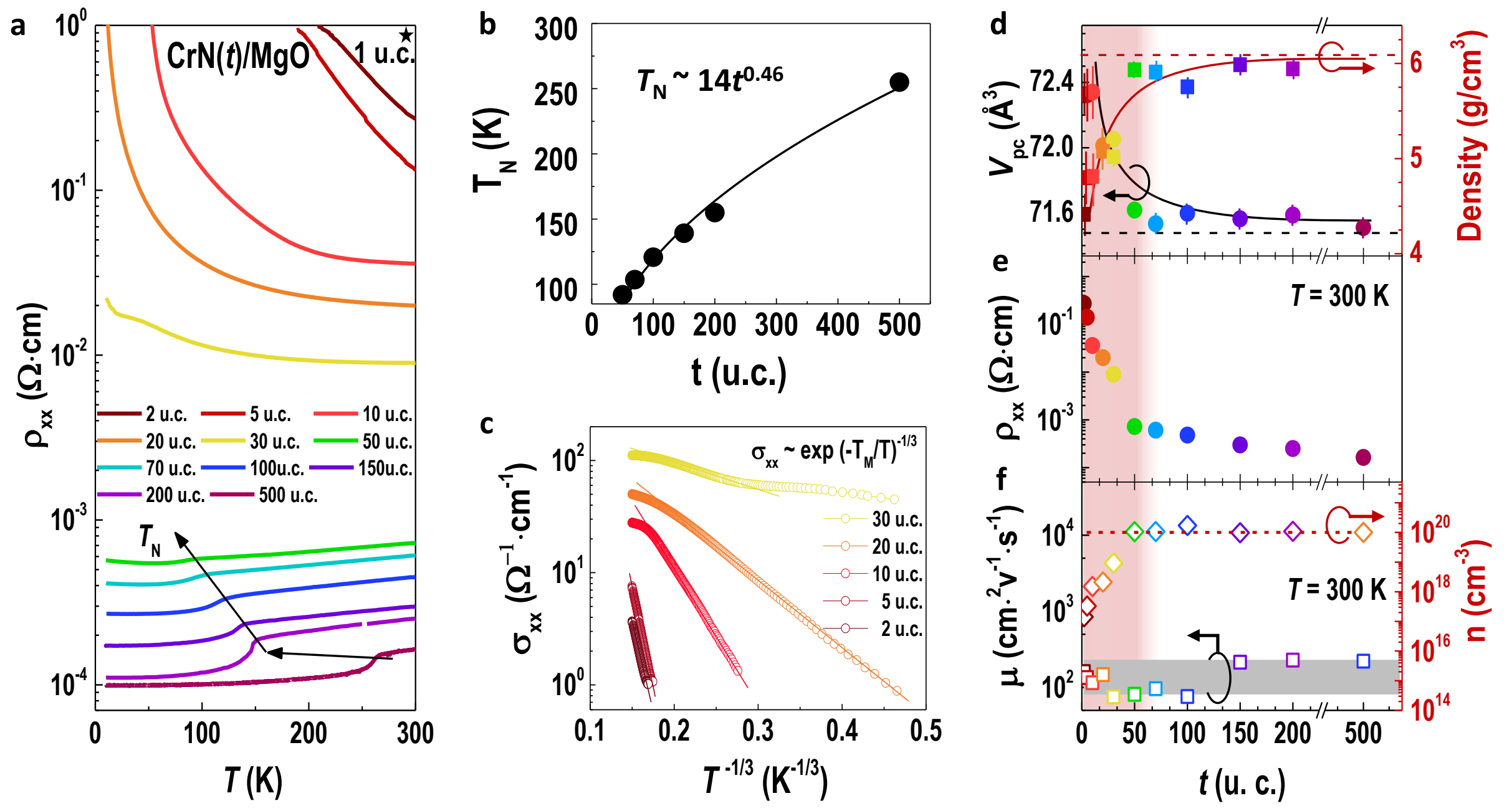

**Figure 2. Thickness-dependent transport properties of CrN films.** (a) $\rho_{xx}$-$T$ curves of CrN films with different thicknesses ($t$) ranging from 1 to 500 u.c. (b) $T_N$ of CrN films reduce progressively as a function of layer thickness ($t$). The solid line is the result of finite-size scaling relation with $T_N$ is proportional to the $B(t)^{\lambda}$, where extrapolated B = 14 ± 0.5 and the shift exponent for the finite-size scaling factor $\lambda = 0.46 \pm 0.05$. (c) We plot the conductivities ($\sigma_{xx}$) of insulating CrN films as a function of $T^{-1/3}$, yielding a linear relationship of $\sigma_{xx} \sim \exp(T_0/T)^{-1/3}$, where $T_0$ is the temperature normalization constant. The exponent 1/3 indicates that the CrN ultrathin films transit into the two-dimensional (2D) conduction. (d) Room-temperature pseudocubic lattice volume ($V_{pc}$, left y-axis) and density (righ y-axis), (e) $\rho_{xx}$ at 300 K, and (f) Hall mobility ($\mu$, left y-axis) and carrier density ($n$, right y-axis) versus film thickness. $V_{pc}$ is obtained from both XRD and RSM measurements and the densities are derived from XRR measurements.

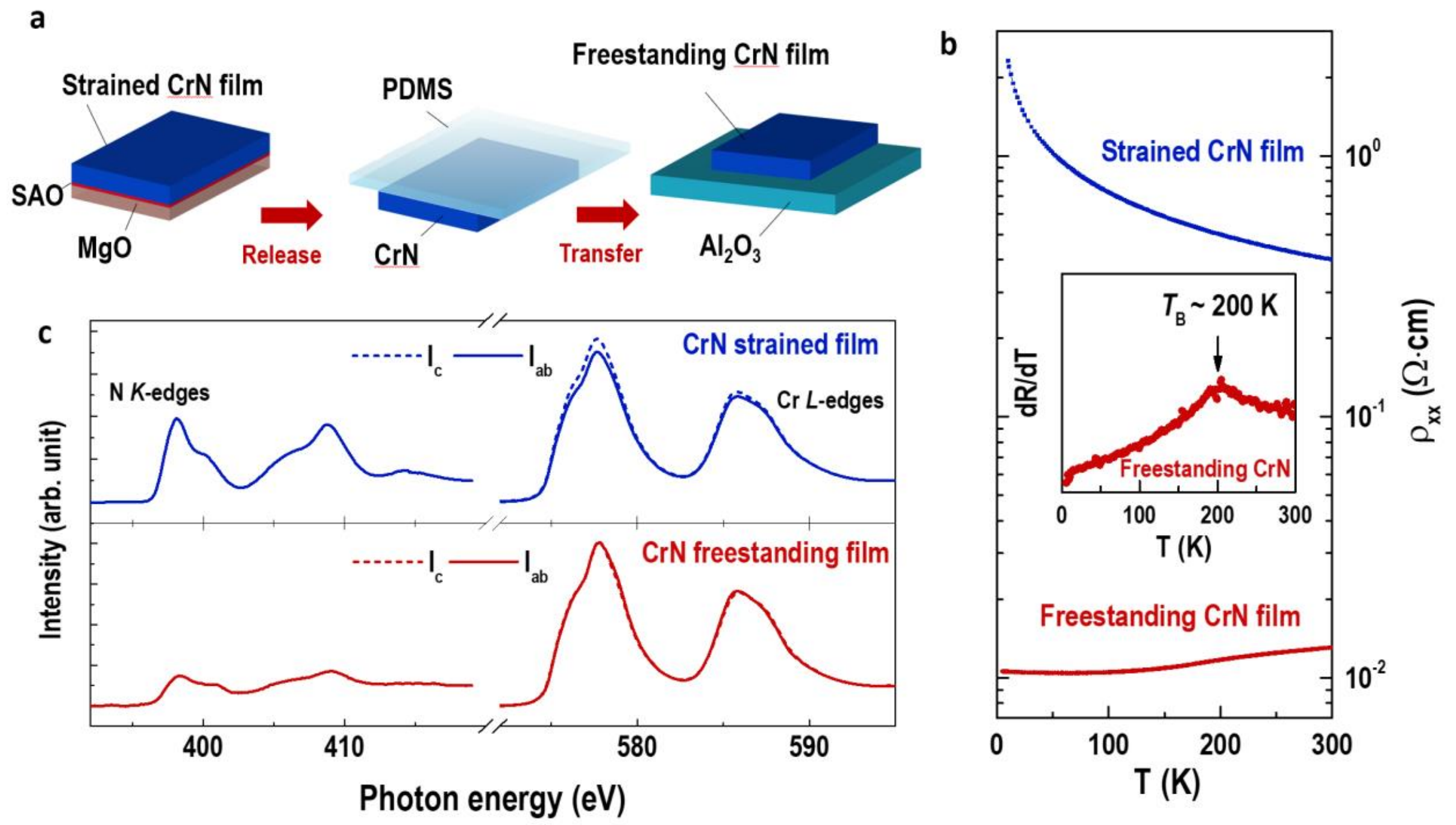

**Figure 3. Fabrication and characterizations of freestanding CrN membranes.** (a) Schematic of the original CrN films grown on a $Sr_3Al_2O_6$ (SAO)-covered MgO substrate, the released CrN film on PDMS support, and the transferred CrN film on sapphire wafer. (b) $\rho_{xx}$-$T$ curves of strained and freestanding CrN films. The freestanding sample is obtained by peeling of the strained film layer from its substrate. Both samples have an identical thickness of 20 u.c. Inset shows d$R$/d$T$ versus temperature for a freestanding CrN film, indicating an electronic transition at ~200 K. (c) XAS for a strained CrN film and a freestanding CrN film at both N $K$- and Cr $L$-edges. Solid (dashed) lines represent the calculated XAS from the linearly polarized x-ray beams with in-plane (out-of-plane) component parallel to the $d_{x^2-y^2}$ ($d_{3z^2-r^2}$) orbital.

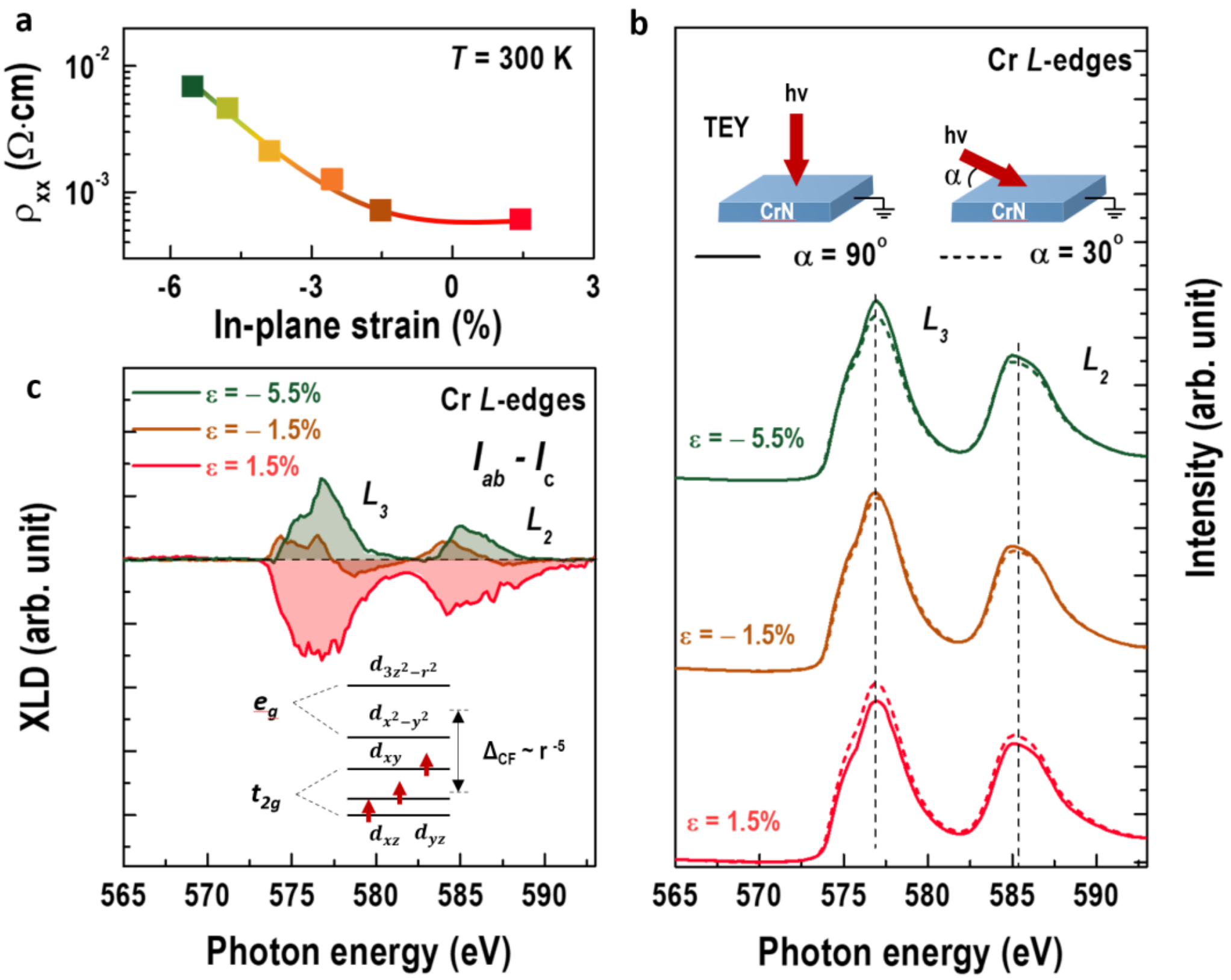

**Figure 4. Strain-dependent resistivity and electronic state of CrN thin films.** (a) Room-temperature $\rho_{xx}$ versus in-plane strain. All CrN films with a thickness of ~ 70 u.c. are epitaxially grown on various substrates. The in-plane strain is calculated from RSM measurements. (b) XAS spectra of Cr *L*-edges for three different CrN films with in-plane strain states of 1.5%, -1.5%, and -5.5%. The XAS measurements were taken by rotating the samples with respect to the direction of incident beam at the angles (α) of 30° (dashed lines) and 90° (solid lines), respectively. Insets are schematic diagrams of the scattering geometry for XAS measurements. (c) XLD spectra of the Cr *L*-edges for three CrN films. XLD is calculated by the normalized intensity, i.e., $(I_{ab}-I_c)$. The strain is clearly induced strongly anisotropic orbital occupancies. Inset is the schematic diagram of Cr: d orbitals in cubic CrN. The crystal field splitting energy ($\Delta_{CF}$) is proportional to the distance ($r$) between two Cr atoms.

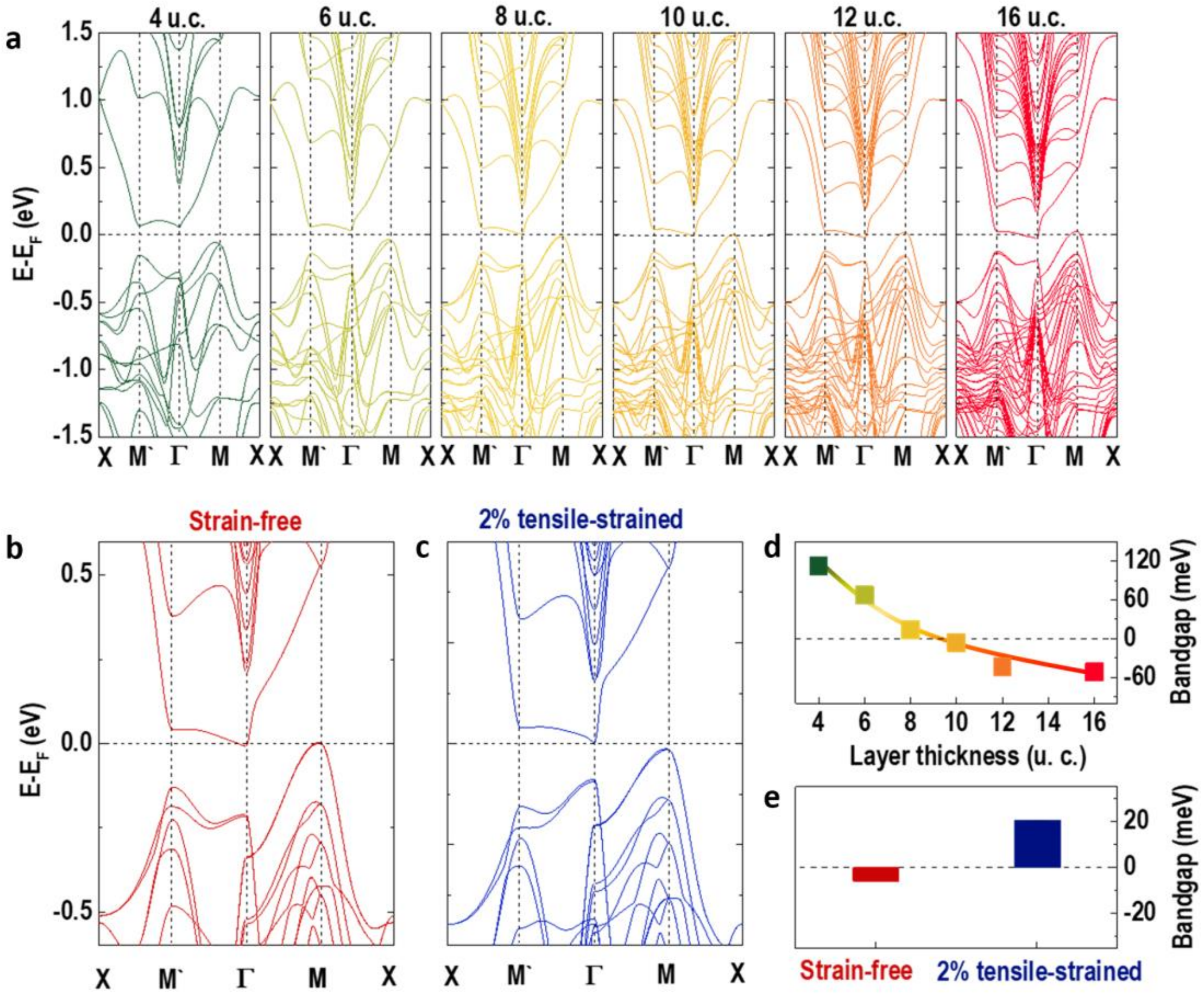

**Figure 5. Thickness- and strain-dependent evolution of band structures of CrN films.** (a) Band structures of CrN films with thickness ranging from 4 to 16 u.c. Comparison of band structures for CrN films with fixed thickness of 10 u.c. between (b) strain-free and (c) tensile strain of 2%. The bandgap of the tensile-strained CrN film is 20 meV, suggesting an insulating state. While the bandgap closes with a relative value of −6.1 meV, demonstrating that the strain-free CrN film transforms into a metal after the removal of tensile strain. (d) Thickness-dependent bandgap of CrN films. The bandgap is defined by the difference between the eigenvalues at the bottom of conduction band and the top of valence band. The insulating films show a positive bandgap, whereas the metallic films exhibit a negative bandgap. (e) A direct comparison of bandgaps between a strain-free CrN film and a CrN film under a tensile strain of 2% with a fixed thickness of 10 u.c. Clearly, the release of tensile strain will trigger an insulator-to-metal phase transition.